\documentclass[11pt]{article}
\usepackage[dvips]{graphics}
\title{Strings, world-sheet covariant quantization and Bohmian mechanics}
\author{Hrvoje Nikoli\'c \\
Theoretical Physics Division, Rudjer Bo\v{s}kovi\'{c} Institute, \\
P.O.B. 180, HR-10002 Zagreb, Croatia \\
{\normalsize e-mail: hrvoje@thphys.irb.hr} \\
\makebox[1in]{} \\
}
\date{\today}
\begin{document}
\maketitle
\begin{abstract}
The covariant canonical method of quantization 
based on the De Donder-Weyl covariant canonical formalism is used 
to formulate a world-sheet covariant quantization of 
bosonic strings. To provide the consistency 
with the standard non-covariant canonical quantization, 
it is necessary to adopt a Bohmian deterministic 
hidden-variable equation of motion. In this way, string theory 
suggests a solution to the problem of measurement
in quantum mechanics. 
\end{abstract}
\vspace*{0.5 cm}

%\pacs{11.25.-w; 04.60.Ds; 03.65.Ta} 
%strings and branes, canonical quantization, foundations of QM (measurement)
%Keywords: Bosonic string; World-sheet covariance; Bohmian mechanics 
%\maketitle

\section{Introduction}

String theory \cite{gsw,polc,zwie} is a theory with an ambition 
to be the theory of everything. However, there is one fundamental 
problem on which, so far, string theory has had nothing new to say.
This is the problem of interpretation of quantum mechanics (QM), 
or in more physical terms, the problem of measurement in QM.
From this point of view, it is widely believed that
it does not matter whether one quantizes a particle, a field, or a
string; the formalism of quantization is, essentialy, always the same,
so the interpretation adopted, say, for particles, should work equally well
(or badly) for fields or strings.
Although the recent progress in understanding the phenomenon of 
decoherence shed much light on the problem of measurement in QM, 
this problem is still considered unsolved \cite{schl}. 
In this paper, however, we argue 
that string theory offers a {\em new} insight into the problem 
of interpretation/measurement in QM, an insight that cannot be inferred 
from the quantization of a particle. Since strings, unlike particles, 
are extended objects, the requirement of world-sheet covariance 
leads to a non-trivial relation between the $\sigma^0$-dependence 
and the $\sigma^1$-dependence of the string coordinates
$X^{\alpha}(\sigma^0,\sigma^1)$.
In order to preserve the world-sheet covariance at the quantum level, 
we argue that the classical covariant De Donder-Weyl canonical formalism 
(see e.g. \cite{kas,rieth} and references therein) 
might be a good starting point.
The appropriate quantum formalism is developed in
\cite{nikolepjc} for fields. In particular, the formalism attributes a new 
status to the Bohmian deterministic hidden-variable
interpretation of QM
\cite{bohm,bohmrep1,bohmrep2,holrep,holbook,nikolfpl1,nikolfpl2},
because, in \cite{nikolepjc}, the Bohmian equations of motion for fields
are {\em derived} from the requirement of spacetime covariance.
By replacing the requirement of spacetime covariance for fields
with that of world-sheet covariance for strings,
in this paper we observe that a completely analogous argument
leads to the Bohmian formulation of quantum strings.
(In the Bohmian interpretation, the quantum string coordinates 
$X^{\alpha}(\sigma^0,\sigma^1)$ evolve in a deterministic manner 
even when they are not measured.)
Thus, in contrast to particle physics 
where the Bohmian deterministic 
interpretation is just one of many interpretations of QM, 
we argue that in string theory the Bohmian interpretation emerges naturally 
from the requirement of world-sheet covariance. 

To further motivate the analysis presented in subsequent sections,
it is worthwhile to explain the conceptual difference between the 
physical meaning of the results obtained in \cite{nikolepjc}  
and that of the present paper. For that purpose, we need 
to recapitulate the concepts of particles, fields and strings in a 
somewhat wider context. In non-string theories, quantum fields 
are often viewed in {\em two} different ways. The prevailing point of view 
among ``hard-core" field theorists 
is that fields are the only fundamental objects, while particles
are merely emergent objects that sometimes even cannot be well defined 
(see e.g. \cite{ful,bd,dav,zeh}).
On the other hand, particle-physics phenomenologists are more willing 
to view pointlike particles as the fundamental objects, while 
fields are often viewed among them merely as a calculational tool convenient 
for treating interactions in which the number of particles changes.
Indeed, there exists an alternative string-inspired  
particle-scattering formalism that completely avoids any referring to fields 
\cite{schub}. In string theory, the situation is similar, but with a 
difference consisting in the fact that {\em most} of the work in string theory 
is done without referring to string-field theory. Moreover, there 
are indications that string-field theory might {\em not} be the 
correct way to treat string interactions \cite{polc2}. Thus, 
from the string-theory perspective,  
particles might be more fundamental objects than fields. 

In the context of the Bohmian hidden-variable interpretation of QM,
the field-or-particle dilemma is even sharper than in the 
conventional interpretation. Should the Bohmian interpretation be 
applied to particles, to fields, or to both? Since the conventional 
probabilistic interpretation cannot be applied to the relativistic 
Klein-Gordon equation, the Bohmian deterministic interpretation
of relativistic quantum particles
might be a natural choice with interesting measurable 
predictions \cite{nikolno,nikolfpl3}. 
However, a {\em derivation}
of the Bohmian interpretation from the requirement of relativistic 
covariance based on the De Donder-Weyl formalism \cite{nikolepjc}
works for fields, but not for particles. Thus, if particles 
are more fundamental objects than fields, then the results of 
\cite{nikolepjc} might be physically irrelevant and we still 
cannot {\em derive} the Bohmian interpretation. 
However, now comes string theory that saves the situation.
If particles are more fundamental than fields, but if 
they are not really pointlike, but extended objects as in 
string theory, then the results of \cite{nikolepjc} 
{\em can} be applied. In this case, the Bohmian interpretation 
of strings can be derived from the requirement of world-sheet covariance, 
while the resulting string theory in a pointlike-particle limit reduces 
to the Bohmian interpretation of relativistic quantum particles.      

The classical De Donder-Weyl formalism for bosonic strings 
is presented in Sec.~\ref{CLAS}, while the corresponding 
quantum theory of bosonic strings is formulated in Sec.~\ref{QUANT}. 
The case of supersymmetric strings is still beyond 
our current technical achievements.

\section{Classical De Donder-Weyl formalism for bosonic strings} 
\label{CLAS}

In order to have a notation similar to that in \cite{nikolepjc},
let the letters $\alpha,\beta=0,1,\cdots,D-1$ denote 
the target spacetime indices, and let the letters $\mu,\nu=0,1$ 
denote the world-sheet indices. The signature of the 
spacetime metric is chosen to be $(+,-,\ldots,-)$. Similarly,
on a flat world-sheet we have $\eta^{00}=-\eta^{11}=1$.
We also use the notation $\sigma\equiv(\sigma^0,\sigma^1)$. 
With this notation, the action of a bosonic string is
\begin{equation}\label{action}
A=\int d^2\sigma\,{\cal L} ,
\end{equation}  
where
\begin{equation}
{\cal L}=-\frac{1}{2}|h|^{1/2}h^{\mu\nu}\eta_{\alpha\beta}
(\partial_{\mu}X^{\alpha})(\partial_{\nu}X^{\beta}) 
\end{equation}
is the Lagrangian density. Here $\eta_{\alpha\beta}$ is 
a flat Minkowski metric in $D$ dimensions, 
$h^{\mu\nu}(\sigma)$ is an arbitrary metric on the string 
world-sheet, and $h$ is the determinant of $h_{\mu\nu}$.
The spacetime and world-sheet indices 
are raised (lowered) by $\eta^{\alpha\beta}$ ($\eta_{\alpha\beta}$)
and $h^{\mu\nu}$ ($h_{\mu\nu}$), respectively.
By requiring that the variation of (\ref{action}) with respect  
to $h^{\mu\nu}$ should vanish, one obtains that $h_{\mu\nu}$ must be 
proportional to the induced metric on the world-sheet \cite{zwie}, 
i.e.
\begin{equation}\label{constr}
h_{\mu\nu}(\sigma)=f(\sigma)(\partial_{\mu}X^{\alpha})
(\partial_{\nu}X_{\alpha}) ,
\end{equation}
where $f(\sigma)$ is an arbitrary positive-valued function.

The canonical momentum world-sheet {\em vector density} is 
defined as
\begin{equation}\label{P}
{\cal P}_{\alpha}^{\mu}=
\frac{\partial{\cal L}}{\partial(\partial_{\mu}X^{\alpha})}=
-|h|^{1/2}\partial^{\mu}X_{\alpha} .
\end{equation}
The covariant De Donder-Weyl Hamiltonian density is given by the 
Legendre transform
\begin{eqnarray}\label{H}
{\cal H} & = & {\cal P}_{\alpha}^{\mu}\partial_{\mu}X^{\alpha}
               -{\cal L} \nonumber \\
         & = & -\frac{1}{2} \frac{h_{\mu\nu}}{|h|^{1/2}} \eta^{\alpha\beta}
               {\cal P}_{\alpha}^{\mu}{\cal P}_{\beta}^{\nu} .
\end{eqnarray}
When (\ref{P}) is satisfied, then ${\cal H}={\cal L}$.
The covariant Hamilton equations of motion are
\begin{equation}\label{em}
\partial_{\mu}X^{\alpha}=\frac{\partial{\cal H}}
{\partial{\cal P}_{\alpha}^{\mu}}, \;\;\;\;
\partial_{\mu}{\cal P}_{\alpha}^{\mu}=-\frac{\partial{\cal H}}
{\partial X^{\alpha}} .
\end{equation}
Using (\ref{H}), we see that
the first equation in (\ref{em}) is equivalent to (\ref{P}). 
Since ${\cal H}$ in (\ref{H}) does not depend on $X^{\alpha}$, 
the second equation in (\ref{em}) leads to the covariant string-wave equation 
\begin{equation}\label{emw}
\partial_{\mu}(|h|^{1/2}\partial^{\mu}X_{\alpha})=0.
\end{equation}
Thus, the classical De Donder-Weyl covariant canonical formalism is equivalent 
to the classical Lagrangian formalism which also leads to the covariant equation 
of motion (\ref{em}). Similarly, it is also equivalent to the
ordinary non-covariant Hamilton formalism, in which the Hamiltonian 
is defined such that only $\mu=0$ contributes in the first line of
(\ref{H}).  

The next step is to introduce the covariant De Donder-Weyl Hamilton-Jacobi 
formalism. We introduce a vector-density function 
${\cal S}^{\mu}(X(\sigma),\sigma)$ that satisfies the De Donder-Weyl
Hamilton-Jacobi equation
\begin{equation}\label{HJ1}
{\cal H}+\partial_{\mu}{\cal S}^{\mu}=0 . 
\end{equation}
Here ${\cal H}$ is given by (\ref{H}) with the replacement 
\begin{equation}
{\cal P}_{\alpha}^{\mu} \rightarrow
\frac{\partial {\cal S}^{\mu}}{\partial X^{\alpha}} .
\end{equation}
The partial derivative $\partial_{\mu}$
acts only on the second argument of ${\cal S}^{\mu}(X(\sigma),\sigma)$.
The corresponding total derivative is given by
\begin{equation}\label{deriv}
d_{\mu}=\partial_{\mu}+(\partial_{\mu}X^{\alpha})
\frac{\partial}{\partial X^{\alpha}} .
\end{equation} 
For a given solution ${\cal S}^{\mu}(X,\sigma)$ of
the De Donder-Weyl Hamilton-Jacobi equation, the $\sigma$-dependence 
of $X^{\alpha}(\sigma)$ is determined by the equation od motion 
\begin{equation}\label{emhj1}
-|h|^{1/2}\partial^{\mu}X_{\alpha}=\frac{\partial {\cal S}^{\mu}}
{\partial X^{\alpha}} .
\end{equation}
 
The classical De Donder-Weyl Hamilton-Jacobi formalism above 
has a manifest world-sheet covariance. We would like 
to construct an analogous quantum formalism with a 
manifest world-sheet covariance at the quantum level. 
It is already known how to construct the
quantum formalism that corresponds to the  
ordinary non-covariant Hamilton-Jacobi formalism: by using 
quantum mechanics represented by the Schr\"odinger equation.
Thus, the first step 
towards quantization based on the covariant De Donder-Weyl Hamilton-Jacobi
formalism is to explain how the ordinary non-covariant 
Hamilton-Jacobi formalism can be obtained from the covariant one. 
Choosing $h_{\mu\nu}=\eta_{\mu\nu}$, (\ref{HJ1}) can be written in an 
explicit form
\begin{equation}\label{HJ2}
-\frac{1}{2} \frac{\partial {\cal S}^0}{\partial X^{\alpha}}
\frac{\partial {\cal S}^0}{\partial X_{\alpha}} +
\frac{1}{2} \frac{\partial {\cal S}^1}{\partial X^{\alpha}}    
\frac{\partial {\cal S}^1}{\partial X_{\alpha}}
+\partial_0{\cal S}^{0}+\partial_1{\cal S}^{1}=0 .
\end{equation} 
Using (\ref{deriv}) and (\ref{emhj1}), the last term can be written
as
\begin{equation}
\partial_1{\cal S}^{1}=d_1{\cal S}^{1}-
(\partial_1 X^{\alpha})(\partial_1 X_{\alpha}) .
\end{equation}
Similarly, the second term in (\ref{HJ2}) can be written as
$(1/2)(\partial_1 X^{\alpha})(\partial_1 X_{\alpha})$. 
Now we introduce the quantity
\begin{equation}\label{S}
S=\int d\sigma^1 {\cal S}^0 ,
\end{equation}
so that
\begin{equation}
\frac{\partial{\cal S}^0(X(\sigma),\sigma)}{\partial X^{\alpha}(\sigma)}=
\frac{\delta S([X(\sigma^0,\sigma^1)],\sigma^0)}
{\delta X^{\alpha}(\sigma^1;\sigma^0)} ,
\end{equation}
where
\begin{equation}\label{funcd}
\frac{\delta}{\delta X^{\alpha}(\sigma^1;\sigma^0)} \equiv
\left.
\frac{\delta}{\delta X^{\alpha}(\sigma^1)} 
\right|_{X(\sigma^1)=X(\sigma)}
\end{equation}
is the functional derivative. Thus, by integrating 
(\ref{HJ2}) over $d\sigma^1$, we obtain the ordinary non-covariant 
Hamilton-Jacobi equation
\begin{equation}\label{HJ3}
H+\partial_0 S=0 ,
\end{equation}
where
\begin{eqnarray}\label{hamHJ3}
H & = & 
- \int d\sigma^1 \left[ \frac{1}{2} 
\frac{\delta S}{\delta X^{\alpha}(\sigma^1;\sigma^0)}
\frac{\delta S}{\delta X_{\alpha}(\sigma^1;\sigma^0)} 
\right. \nonumber \\
& & \left.
+\frac{1}{2} (\partial_1 X^{\alpha})(\partial_1 X_{\alpha}) \right] 
\end{eqnarray}
is written for the $\sigma^0$-dependent string coordinate
$X^{\alpha}(\sigma^0,\sigma^1)$. 
The integral of a total derivative
$\int d\sigma^1\, d_1{\cal S}^{1}$ is ignored because it is a 
constant without any physical significance.
The $\sigma^0$-evolution of
$X^{\alpha}(\sigma^0,\sigma^1)$ is given by 
\begin{equation}\label{emhj2}
-\partial^0 X_{\alpha}(\sigma^0,\sigma^1)=
\frac{\delta S}{\delta X^{\alpha}(\sigma^1;\sigma^0)} ,
\end{equation}
which is a consequence of the $\mu=0$ component of (\ref{emhj1}).
The covariant constraint (\ref{constr}) implies the 
non-covariant Hamiltonian constraint $H=0$ \cite{zwie}.

To anticipate the implications to the quantum case, here 
it is crucial to observe the following. 
First, to derive (\ref{HJ3}) from (\ref{HJ1}), it was 
necessary to use the $\mu=1$ component of (\ref{emhj1}).
Second, if the world-sheet covariance is required, then 
the validity of the $\mu=1$ component of (\ref{emhj1}) 
also implies the validity of the $\mu=0$ component of (\ref{emhj1}).
Third, the validity of the $\mu=0$ component of (\ref{emhj1})
implies the classical determinism incoded in (\ref{emhj2}).  
Thus, the determinism in classical string theory can be 
derived from the world-sheet covariance and 
the requirement that the covariant Hamilton-Jacobi equation 
(\ref{HJ1}) and the non-covariant one (\ref{HJ3})
should be {\em both} valid. As we shall see in the next section, 
a similar argument leads to a derivation of the Bohmian deterministic
hidden-variable formulation of quantum strings.    

\section{Quantization and Bohmian mechanics} 
\label{QUANT}

How to quantize strings such that the world-sheet covariance is manifest?
The standard method is the path-integral quantization
based on calculating the generating functional
$Z=\int[dX][dh]\exp{(iA/\hbar)}$.
(To avoid an anomaly, one must fix $D=26$ \cite{poly}.)
This method 
is useful for calculating Green functions and scattering amplitudes. 
%from $\sigma^0\rightarrow -\infty$ to $\sigma^0\rightarrow \infty$.
Although this is usually sufficient for calculating quantities 
that are measured in practice, there are also 
quantities that can be measured in principle  
but cannot be calculated in a covariant way from $Z$.
In particular, the generating functional $Z$ does not 
describe a quantum state at a given time. 
Thus, certain physical information is not described
by the path-integral quantization. 
In order to obtain such information, one can try to use 
the $\sigma^0$-dependent
quantum states $\Psi([X(\sigma^1)],\sigma^0)$ that satisfy the
functional Schr\"odinger equation
\begin{equation}\label{sch}
\hat{H}\Psi=i\hbar\partial_0 \Psi ,
\end{equation}
where
\begin{eqnarray}\label{ham}
\hat{H} & = & 
-\int d\sigma^1 \left[ \frac{-\hbar^2}{2}
\frac{\delta}{\delta X^{\alpha}(\sigma^1)}
\frac{\delta}{\delta X_{\alpha}(\sigma^1)} 
\right. \nonumber \\
& & \left.
+\frac{1}{2} (\partial_1 X^{\alpha})(\partial_1 X_{\alpha}) \right] .
\end{eqnarray}
However, not all states satisfying (\ref{sch}) are physical. In particular,
physical states satisfy the Hamiltonian constraint 
\begin{equation}\label{sch0}
(\hat{H}+a)\Psi=0 ,
\end{equation}
where $a$ originates from a constant that can be added to the 
action (\ref{action}) without changing classical
properties of strings. A more common view of this constant 
is in terms of an operator-ordering constant that 
can be fixed uniquely \cite{gsw,polc,zwie}. 
The discussion of the value of $a$, as well as the discussion of 
other requirements on physical states related to 
the requirement of target spacetime covariance, are beyond the 
scope of the present paper.
We only note that (\ref{sch0}) and (\ref{sch}) imply that
all physical states have the same trivial dependence on $\sigma^0$, 
i.e. that 
$\Psi([X(\sigma^1)],\sigma^0)=\Psi[X(\sigma^1)] e^{ia\sigma^0/\hbar}$. 

To write (\ref{sch}) and (\ref{ham}), one has to fix
a special world-sheet coordinate $\sigma^0$.
However, any such choice breaks
the world-sheet covariance. To solve this problem, we
want to find a covariant substitute for the Schr\"odinger equation
(\ref{sch}). The similarity of the Schr\"odinger equation
(\ref{sch}) to the non-covariant Hamilton-Jacobi equation
(\ref{HJ3}) suggests that a covariant substitute for
(\ref{sch}) might be an equation similar to the
covariant De Donder-Weyl Hamilton-Jacobi equation (\ref{HJ1}).
Indeed, the general method of quantization based on the
De Donder-Weyl Hamilton-Jacobi equation is developed in
\cite{nikolepjc}. (For a different method, with problems
discussed in \cite{nikolepjc}, see also \cite{kan,hip}.)
Here we apply the general results of \cite{nikolepjc}
to the case of bosonic strings.

The first step is to write 
\begin{equation}\label{psiRS}
\Psi=Re^{iS/\hbar} ,
\end{equation}
where $R$ and $S$ are real functionals. One finds that the complex 
equation (\ref{sch}) is equivalent to a set of two real 
equations
\begin{eqnarray}\label{HJq0}
& -\displaystyle\int d\sigma^1 \left[ \frac{1}{2}
\frac{\delta S}{\delta X^{\alpha}(\sigma^1)}
\frac{\delta S}{\delta X_{\alpha}(\sigma^1)}
+\frac{1}{2} (\partial_1 X^{\alpha})(\partial_1 X_{\alpha}) -{\cal Q} \right] &
\nonumber \\ 
& + \partial_0 S =0 , &
\end{eqnarray}
\begin{equation}\label{Rq0}
-\int d\sigma^1 \left[ \frac{1}{2}
\frac{\delta R}{\delta X^{\alpha}(\sigma^1)}
\frac{\delta S}{\delta X_{\alpha}(\sigma^1)} -{\cal J} \right]
+ \partial_0 R =0 ,
\end{equation}
where
\begin{eqnarray}\label{QJ}
& {\cal Q}=\displaystyle\frac{\hbar^2}{2R}  \frac{\delta^2 R}
{\delta X^{\alpha}(\sigma^1) \delta X_{\alpha}(\sigma^1)} , & \nonumber \\
& {\cal J}=-\displaystyle\frac{R}{2} \frac{\delta^2 S}
{\delta X^{\alpha}(\sigma^1) \delta X_{\alpha}(\sigma^1)} . &
\end{eqnarray}
We see that (\ref{HJq0}) is very similar to (\ref{HJ3}) with (\ref{hamHJ3}),
differing from it only in containing the additional quantum ${\cal Q}$-term. 

Now, following \cite{nikolepjc}, we replace the classical
De Donder-Weyl Hamilton-Jacobi equation (\ref{HJ1}) with the 
quantum one
\begin{equation}\label{HJq}
-\frac{1}{2} \frac{h_{\mu\nu}}{|h|^{1/2}} \eta^{\alpha\beta}
  \frac{d {\cal S}^{\mu}}{d X^{\alpha}}
  \frac{d {\cal S}^{\nu}}{d X^{\beta}} 
+{\cal Q}
+\partial_{\mu}{\cal S}^{\mu}=0 .
\end{equation}
Here ${\cal S}^{\mu}([X],\sigma)$ is a functional of $X(\sigma)$ 
and a function of $\sigma$, which incorporates quantum nonlocalities 
in a covariant manner. The derivative $d/d X^{\alpha}$ is a 
generalization of the derivative $\partial/\partial X^{\alpha}$, 
such that the action of the derivative on nonlocal functionals 
is well defined \cite{nikolepjc}. The quantum potential ${\cal Q}$
is defined as in (\ref{QJ}), but with the replacement
$\delta/\delta X^{\alpha} \rightarrow \delta/\delta_C X^{\alpha}$.
The derivative
$\delta/\delta_C X^{\alpha}$ is a covariant version of the derivative 
(\ref{funcd}). The label $C$ denotes a curve on the world-sheet 
that generalizes the curve $\sigma^0={\rm constant}$ in (\ref{funcd}).
The foliation of the world-sheet into curves $C$ is induced 
by the dynamical vector density ${\cal R}^{\mu}([X],\sigma)$;
the curves are defined by requiring that ${\cal R}^{\mu}$ should be 
orthogonal to the curves at each point $\sigma$. The vector density
${\cal R}^{\mu}$ satisfies the dynamical equation of motion
\begin{equation}\label{Rq}
-\frac{1}{2} \frac{h_{\mu\nu}}{|h|^{1/2}} \eta^{\alpha\beta}
  \frac{d {\cal R}^{\mu}}{d X^{\alpha}}
  \frac{d {\cal S}^{\nu}}{d X^{\beta}}
+{\cal J}
+\partial_{\mu}{\cal R}^{\mu}=0 ,
\end{equation}
where ${\cal J}$ is defined as in (\ref{QJ}) with
$\delta/\delta X^{\alpha} \rightarrow \delta/\delta_C X^{\alpha}$. 
The functionals $R$ and $S$ are defined in a covariant way as
\begin{equation}\label{RS}
R=\int_C d\Sigma_{\mu}R^{\mu} , \;\;\;\;
S=\int_C d\Sigma_{\mu}S^{\mu} , 
\end{equation}
where $R^{\mu}=|h|^{-1/2}{\cal R}^{\mu}$ and 
$S^{\mu}=|h|^{-1/2}{\cal S}^{\mu}$ transform as vectors. In the measure 
$d\Sigma_{\mu}=dl\, n_{\mu}$, $dl$ is an element of the invariant length 
of $C$, while $n_{\mu}$ is a unit vector orthogonal to $C$. 
Note that the second equation in (\ref{RS}) is a covariant version 
of (\ref{S}).    
The functionals $R$ and $S$ in (\ref{RS}) define the
wave functional $\Psi$ as in (\ref{psiRS}).

From the covariant formalism above, the non-covariant
Schr\"odinger equation (\ref{sch}) can be derived as a special case.
Assume that ${\cal R}^{\mu}=({\cal R}^0,0)$, that ${\cal S}^1$ is 
a local functional, and that ${\cal R}^0$ and ${\cal S}^0$
are functionals local in the coordinate $\sigma^0$ 
(see \cite{nikolepjc} for the precise definitions of these notions 
of locality!). Then, similarly as in the classical case,
by choosing $h_{\mu\nu}=\eta_{\mu\nu}$ and 
integrating equations (\ref{HJq}) and (\ref{Rq})
over $d\sigma^1$, 
one recovers equations (\ref{HJq0}) and (\ref{Rq0}), which, in turn, 
are equivalent to the Schr\"odinger equation (\ref{sch}).
(The constant originating from the integral
$\int d\sigma^1\, d_1{\cal S}^{1}$ can be absorbed into the 
constant $a$.)
Just as in the classical case, to obtain (\ref{HJq0}) from 
(\ref{HJq}), it is necessary to assume that a quantum analog 
of the $\mu=1$ component of (\ref{emhj1}) is valid. 
The covariance then implies that the $\mu=0$ component is also 
valid, so we have a covariant quantum relation  
\begin{equation}\label{emhj1q}
-|h|^{1/2}\partial^{\mu}X_{\alpha}=\frac{d {\cal S}^{\mu}}
{d X^{\alpha}} .
\end{equation}
The $\mu=0$ component of (\ref{emhj1q})
implies that the non-covariant Schr\"odinger equation (\ref{sch})
should be supplemented with (\ref{emhj2}). In the quantum context, 
equation (\ref{emhj2}) is nothing but the Bohmian deterministic 
equation of motion for the $\sigma^0$-dependent hidden variable 
$X^{\alpha}(\sigma^0,\sigma^1)$. Indeed, by analogy with 
the Bohmian interpretation of particles and fields
\cite{bohm,bohmrep1,bohmrep2,holrep,holbook,nikolfpl1,nikolfpl2},
equation (\ref{emhj2})
could have been postulated immediately after writing 
(\ref{sch}), as an equation that provides a consistent 
Bohmian deterministic
hidden-variable interpretation of quantum strings. 
In this interpretation, the wave function is a physical 
object which does not ``collapse" during measurements. 
The nonlocality incoded in the wave function reflects in a 
nonlocal quantum potential ${\cal Q}$, which provides 
nonlocalities needed for a hidden-variable theory to be 
consistent with the Bell theorem. In the deterministic Bohmian 
interpretation, all quantum uncertainties are an artefact
of the ignorance of the actual initial conditions 
$X^{\alpha}(\sigma^1)$ at some initial $\sigma^0$. 
For more details on this interpretation, we refer the reader 
to the seminal work \cite{bohm} and reviews 
\cite{bohmrep1,bohmrep2,holrep,holbook}.   
Here, however, the crucial equation of
the Bohmian interpretation, Eq.~(\ref{emhj2}), 
is {\em not postulated}
for interpretational purposes, but {\em derived} from the 
requirement of world-sheet covariance! To be more 
precise, we stress that the covariant quantum
equations (\ref{HJq}) and (\ref{Rq}) by themselves do not imply 
the determinism covariantly incoded in (\ref{emhj1q}). 
Instead, the need for the determinism incoded in (\ref{emhj1q})
emerges from the 
requirement that these covariant equations should be compatible 
with standard non-covariant canonical quantum equations. 

Eqs. (\ref{HJq}) and (\ref{emhj1q}) imply a quantum version of
(\ref{emw}), namely
\begin{equation}\label{emwq}
\partial_{\mu}(|h|^{1/2}\partial^{\mu}X_{\alpha})
+\frac{d{\cal Q}}{dX^{\alpha}}=0.
\end{equation}
The covariant quantum constraint takes the same form as 
the classical one (\ref{constr}). 
(Note that (\ref{constr}) would be meaningless in the conventional 
interpretation of the Schr\"odinger picture
that does not attribute a definite dependence 
on $\sigma^0$ to $X^{\alpha}$.)
The non-covariant 
quantum constraint (\ref{sch0}) can be derived from the 
covariant one in a similar way as (\ref{HJq0}) and (\ref{Rq0})
are derived from (\ref{HJq}) and (\ref{Rq}), 
provided, in addition, that a constant is added 
to the action and that ${\cal R}^0$ does not explicitly depend on 
$\sigma^0$.

We also note that the constraint (\ref{sch0}) in the pointlike-particle 
limit reduces to the massless Klein-Gordon equation, provided that 
$a=0$ in the pointlike-particle limit.
(A heuristic way to obtain $a=0$ in the pointlike-particle limit
is to recall \cite{gsw,polc,zwie} that, 
in bosonic string theory, $a$ turns out
to be proportional to $\sum_{n=0}^{\infty}n$, which leads to a finite
value after the analytic continuation of the zeta function. 
Since $n$ represents the wave-mode number of a string, only $n=0$
contributes in the pointlike-particle limit, which leads
to $a=0$.) 
The Bohmian equation of motion (\ref{emhj2}) leads to 
\begin{equation}\label{bohmpart}
\frac{dX_{\alpha}}{ds}=-\frac{\partial S}{\partial X^{\alpha}}
\end{equation}
in the pointlike-particle limit, where $s\equiv\sigma^0$. 
This can be viewed as a stringy derivation of the 
relativistic-covariant Bohmian interpretation of the 
massless Klein-Gordon equation, which, in turn, leads to 
interesting measurable predictions \cite{nikolfpl3}. 

Of course, string theory can also describe particles with arbitrary 
spin, not by considering the poinlike limit, but by 
considering string states $\Psi$ with fixed quantum numbers 
that determine spin (see e.g. \cite{zwie}). 
By integrating (\ref{emhj2}) over $d\sigma^1$, one obtains 
the Bohmian equation of motion (\ref{bohmpart}) for a particle 
with an arbitrary integer spin. 
If bosonic strings are replaced with superstrings, then
the Bohmian interpretation of half-integer spin particles can also 
be included in the same way. 
However, we do not know yet how to formulate the quantum 
de Donder-Weyl formalism for supersymmetric strings, so we have 
not yet been able to {\em derive} the Bohmian equation of motion for all 
spins from the requirement of world-sheet covariance. Our current
technical achievements allow only to derive the Bohmian equation 
of motion for particles with integer spin.      

To summarize, in this paper we have used a new manifestly 
covariant canonical method of quantization developed in 
\cite{nikolepjc} to quantize bosonic strings in a way that 
provides a manifest world-sheet covariance. This new method 
of quantization, based on the classical De Donder-Weyl covariant 
canonical formalism, is more general than the standard 
non-covariant canonical quantization in the Schr\"odinger picture.
The covariant method of quantization contains the non-covariant 
one as a special case (see also \cite{nikolepjc} for a discussion 
about that point). From the requirement that the covariant method 
of quantization should lead to the standard non-covariant quantization 
without violating covariance, it turns out 
that the quantization method should be supplemented with an 
equation that corresponds to the Bohmian deterministic hidden-variable 
formulation of QM. 
Thus, string theory together with the new quantization method 
proposed in \cite{nikolepjc} offers a new insight into the problem
of interpretation and measurement in QM (an insight that cannot be inferred
from the quantization of a particle) by {\em deriving} the Bohmian 
interpretation from the requirement of world-sheet covariance.    

{\it Acknowledgements}.
This work was supported by the Ministry of Science and Technology of the
Republic of Croatia. 
%under Contract No. 0098002.

\end{document}